\documentclass[sigconf]{acmart}

\usepackage{booktabs} % For formal tables
\usepackage{subfigure}
\usepackage{balance}
\usepackage{multirow}
\usepackage{footmisc}
\usepackage{enumitem}
\usepackage{siunitx,booktabs}
\usepackage{caption}
\usepackage{bbm}  
\usepackage{algpseudocode} 
\usepackage[ruled,linesnumbered]{algorithm2e}
\AtBeginDocument{%
  \providecommand\BibTeX{{%
    \normalfont B\kern-0.5em{\scshape i\kern-0.25em b}\kern-0.8em\TeX}}}

\copyrightyear{2022}
\acmYear{2022}
\setcopyright{acmcopyright}\acmConference[WWW '22]{Proceedings of the ACM Web Conference 2022}{April 25--29, 2022}{Virtual Event, Lyon, France}
\acmBooktitle{Proceedings of the ACM Web Conference 2022 (WWW '22), April 25--29, 2022, Virtual Event, Lyon, France}
\acmPrice{15.00}
\acmDOI{10.1145/3485447.3512011}
\acmISBN{978-1-4503-9096-5/22/04}

\begin{document}
\title{Contrastive Learning with Positive-Negative Frame Mask\\ for Music Representation}

\author{Dong Yao${^1}$, 
	Zhou Zhao${^{1*}}$ , 
	Shengyu Zhang${^{1*}}$, 
	Jieming Zhu${^2}$, 
	Yudong Zhu${^2}$, 
	Rui Zhang${^2}$,\\ 
	Xiuqiang He${^2}$}
\affiliation{%
	\textsuperscript{\rm 1}Zhejiang University, China\\
	\textsuperscript{\rm 2}Huawei Noah's Ark Lab
}
\email{{yaodongai, zhaozhou, sy\_zhang}@zju.edu.cn}
\email{jiemingzhu@ieee.org}
\email{{zhuyudong3, hexiuqiang1}@huawei.com}
\email{rayteam@yeah.net}

\begin{abstract}
Self-supervised learning, especially contrastive learning, has made an outstanding contribution to the development of many deep learning research fields. Recently, researchers in the acoustic signal processing field noticed its success and leveraged contrastive learning for better music representation. Typically, existing approaches maximize the similarity between two distorted audio segments sampled from the same music. In other words, they ensure a semantic agreement at the music level. However, those coarse-grained methods neglect some inessential or noisy elements at the frame level, which may be detrimental to the model to learn the effective representation of music. Towards this end, this paper proposes a novel Positive-nEgative frame mask for Music Representation based on the contrastive learning framework, abbreviated as PEMR. Concretely, PEMR incorporates a Positive-Negative Mask Generation module, which leverages transformer blocks to generate frame masks on Log-Mel spectrogram. We can generate self-augmented negative and positive samples by masking important components or inessential components, respectively. We devise a novel contrastive learning objective to accommodate both self-augmented positives/negatives sampled from the same music. We conduct experiments on four public datasets. The experimental results of two music-related downstream tasks, music classification and cover song identification, demonstrate the generalization ability and transferability of music representation learned by PEMR\footnote{*Corresponding author}.

\end{abstract}

\begin{CCSXML}
<ccs2012>
   <concept>
       <concept_id>10010147.10010257.10010258.10010260</concept_id>
       <concept_desc>Computing methodologies~Unsupervised learning</concept_desc>
       <concept_significance>500</concept_significance>
       </concept>
   <concept>
       <concept_id>10010147.10010257.10010293.10010319</concept_id>
       <concept_desc>Computing methodologies~Learning latent representations</concept_desc>
       <concept_significance>300</concept_significance>
       </concept>
 </ccs2012>
\end{CCSXML}

\ccsdesc[500]{Computing methodologies~Unsupervised learning}
\ccsdesc[300]{Computing methodologies~Learning latent representations}

\keywords{	
Contrastive Learning, Music Representation, Representation Learning, Attention}



\maketitle
\newcommand{\vpara}[1]{\vspace{0.05in}\noindent\textbf{#1 }}
\renewcommand{\shortauthors}{Dong Yao and Zhou Zhao, et al.}
\section{Introduction}

\begin{figure}
	\centering
	\includegraphics[width=0.95\linewidth]{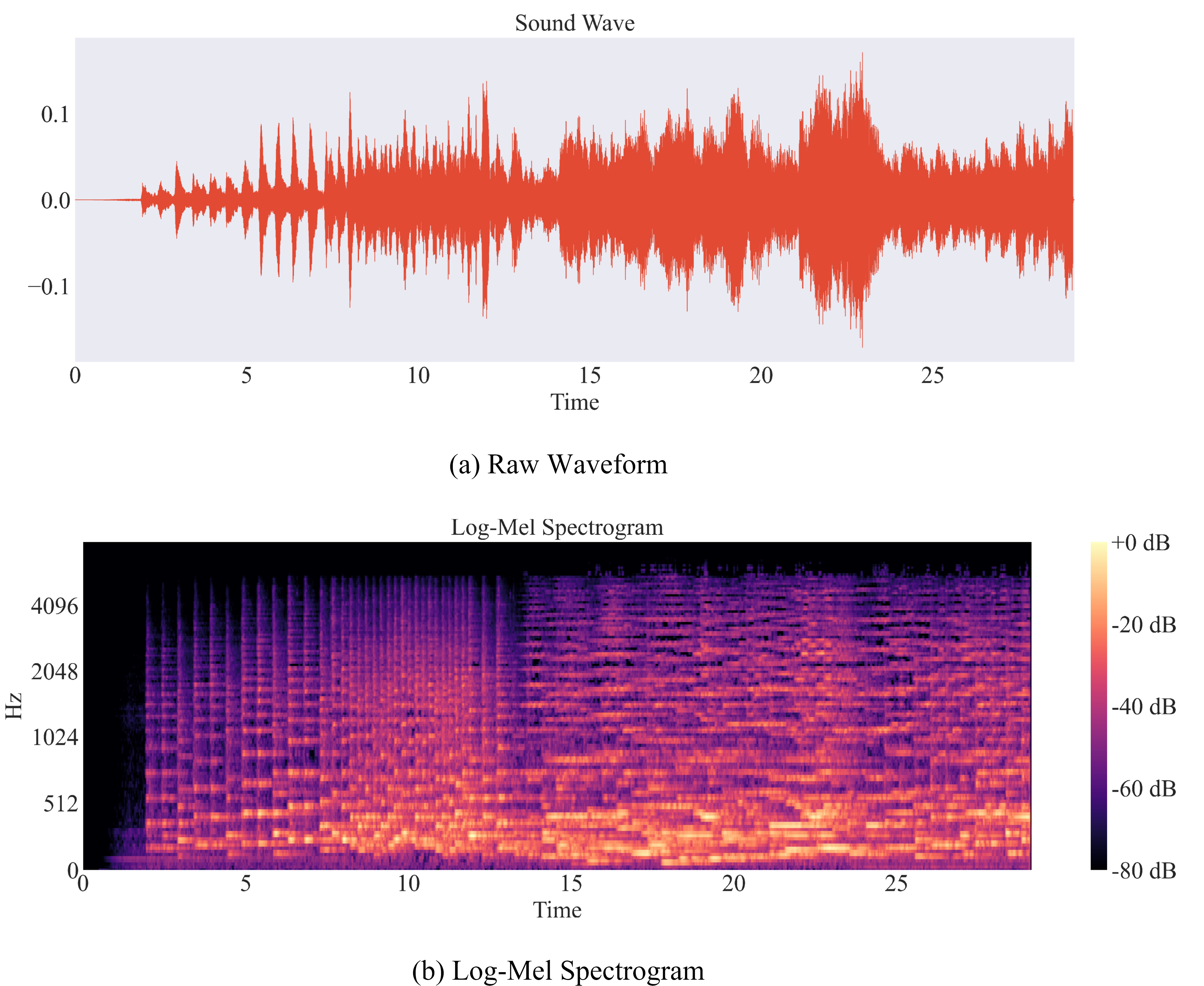}
	\caption{(a) the raw waveform image of a music track, (b) the corresponding Log-Mel spectrogram.}
	\label{fig1}
\end{figure}

Supervised learning has hit a bottleneck. On the one hand, it relies heavily on expensive manual tags and is subject to label errors and false correlations; On the other hand, the amount of labeled data is much smaller than unlabeled data. As a promising alternative, self-supervised learning has drawn massive attention for its data efficiency and generalization ability. Recently, breakthroughs in contrastive learning, such as SimCLR \cite{Chen2020}, MoCo \cite{He2020}, BYOL \cite{Grill2020}, Deep Cluster \cite{Caron2018}, SDCLR \cite{Jiang2021}, shed light on the potential of contrastive learning for learning self-supervised representation. Contrastive learning has increasingly become dominant in self-supervised learning owing to its competitive experimental performance compared with conventional supervised methods.

In Music Information Retrieval (MIR) community, many research-ers have made great efforts to learn effective music representation applied in different music-related tasks, such as music classification \cite{VandenOord2014, Choi2016, Choi2017, Pons2018, Pons2019, Lee2019, Wu2021}, cover song identification \cite{Xu2018, Yu2019, Yesiler2020, Yu2020, Jiang2020}, music generation \cite{Ren2020, Huang2021, Liu2021, Chen2021, Ren2021}. However, most of them learn music representation in a supervised manner. Due to the labeled datasets upon which the supervised learning methods depend being costly and time-consuming, the performance of supervised learning methods will be limited. For that reason, some audio research workers have adopted contrastive learning methods to train neural networks.
The underlying idea of contrastive learning applied in music is to minimize the distance among the audio segments from the same input while minimizing the similarity among the audio segments from the different inputs. CLMR \cite{Spijkervet2021} uses a simple contrastive learning framework for music representation whose encoder directly encodes raw waveforms of songs. Although it performs well in classification downstream task, encoding raw waveforms can hardly encode frequency distribution into the final representation of music. To make the model understand the music in time and frequency domain, unlike CLMR, COLA \cite{Saeed2021} encodes the Log-Mel spectrogram of music so that the time-frequency information can be embedded into music representation. A Mel spectrogram is a spectrogram where the frequencies are converted to the Mel scale. More details about the Log-Mel spectrogram can be found in Section \ref{logMel}. BYOL-A \cite{Niizumi2021}, adopting the network structure of BYOL, which owns an online network and a target network, also opt the Log-Mel spectrogram of the music as model input. Nevertheless, there exists an issue that they encode \textbf{all} frames of the spectrogram into music representation space. That would be harmful to the quality of learned music representation since not all frames impact the music positively. As shown in figure \ref{fig1}, the onset of the track may be silent or directly missing, resulting in the absence of valid content of these starting frames. The quality of music downloaded from the web is uneven. A song may lose its content at the beginning or any other position, while other songs may contain noisy frames. These frames are unimportant parts of the whole music. On the other hand, we argue that each frame has a different status in characterizing music. For example, the drastic parts of rock music are more appropriate than the mild parts when representing the characteristic of a particular song. In other words, the drastic parts are more representative to rock music. Therefore, when learning music representation, we must restrict the non-critical parts of music while augmenting the role of crucial parts.

In order to address the above challenge, we propose to mask some frames within a piece of music with a \textbf{P}ositive-n\textbf{E}gative frame mask for \textbf{M}usic \textbf{R}epresentation. Specifically, a predicting module is designed to catch the correlation of frames. simultaneously, an asymmetrical structure module utilizes the parameters of multi-head attention layers from the transformer encoder to produce the positive-negative mask. The positive mask will erase the existence of inessential frames. Thus, the remaining crucial frames will be encoded and projected into the contrastive learning space to obtain the augmented positive representation. In turn, we can get the counterfactual \cite{Zhang_Yao_Zhao_Chua_Wu_2021} negative representation by adopting the negative mask. Moreover, we design a contrastive learning objective for positive-negative representation pairs. These masks and loss functions can make the model pay more attention to the critical frames and reserve the music's global semantic information while reducing the non-critical frames' adverse effects.

We pre-train the model on several public musical datasets and employ labeled data to train classifiers based on self-supervised representation learned by PEMR. The classifiers achieve the state-of-the-art performance in \textbf{music classification} task. We fine-tune the encoder pre-trained in a dataset on another dataset for classification to evaluate transferability and generalization capability. Besides, we apply the pre-trained encoder in \textbf{cover song identification} task and fine-tune it. We can obtain the more advanced performance for cover song identification by incorporating our pre-trained encoder into the current advanced supervised model. In summary, the contribution of this work is threefold:

\begin{itemize}
	\item We propose to mask some crucial or inessential parts of music so that the inessential parts will be limited and the critical parts will be boosted when learning music representation.  
	
	\item We devise an asymmetrical mask generation module, generating the positive and negative masks for input music and design a contrastive learning loss function. We incorporate them into the contrastive learning framework for learning more effective music representation. 
	
	\item The extensive experimental results show that our learned musical representation achieves state-of-the-art performance on a downstream classification task. Furthermore, our learned representation improves the performance of cover song identification, demonstrating its effectiveness and transferability.
\end{itemize}

\section{Related Work}

\subsection{Contrastive Learning}

To solve the problem of the ever-growing unlabeled data, lots of self-supervised methods \cite{Devlin2019, Fausk2007, Kipf2016, Razavi2019, Goodfellow2020, Caron2018, Caron2020} have been proposed in several areas, especially  Computer Vision, Natural language processing. Since \cite{Hadsell2006} whose approaches contrast positive pairs against negative pairs to learn representation, contrastive learning has attracted a great deal of attention from both academia and the industrial community. Contrastive Predictive Coding \cite{Henaff2019} is an unsupervised objective that learns predictable representation. CMC \cite{Tian2020} is view-agnostic and can scale to any number of views by maximizing mutual information between different views of the same observation for learning representation. MoCo \cite{He2020} views the contrastive learning as a dictionary look-up to build a dynamic queue including samples of the current mini-batch and the previous mini-batch and a moving-averaged encoder. Another method SimCLR \cite{Chen2020}, is a simple framework for contrastive learning without a memory bank. Recently, BYOL \cite{Grill2020} proposed a new architecture for contrastive learning, which consists of online and target networks. They train the networks only with the various augmented views of an identical image without negative pairs. To avoid collapsed solution and minimize the redundancy, \cite{Zbontar2021} contrast samples from features dimension.

Many researchers in the music community have attempted to apply contrastive learning for learning music representation. \cite{Saeed2021} designs a common contrastive model for learning general-purpose audio representation. \cite{Spijkervet2021} also uses SimCLR \cite{Chen2020} framework for pre-training the model. \cite{Niizumi2021} introduces BYOL \cite{Grill2020} for audio and achieves advanced results in various downstream tasks.

\subsection{Masking Strategy in Music Representation}
Masking strategy has played a significant role in the NLP community. The success of BERT \cite{Devlin2019}, which randomly masks some tokens in the input sequence and learns to reconstruct the masked tokens from the output of the transformer encoder, has shown its superiority in learning contextual information among tokens and has attracted the attention of researchers in the audio domain. For example, MusicBERT \cite{Zeng2021} devised a bar-level masking strategy as the pre-training mechanism to understand symbolic music. Mockingjay \cite{Liu2020} is designed to predict the masked frame through jointly conditioning on both past and future contexts. \cite{Zhao2021} proposes two pre-training objectives, including Contiguous Frames Masking (CFM) and Contiguous Channels Masking (CCM),  designed to adapt BERT-like masked reconstruction pre-training to continuous acoustic frame domain.

\section{Proposed Method}

%--------------------------------fig---------------------
\begin{figure*}[t] \begin{center}
		\includegraphics[width=\textwidth]{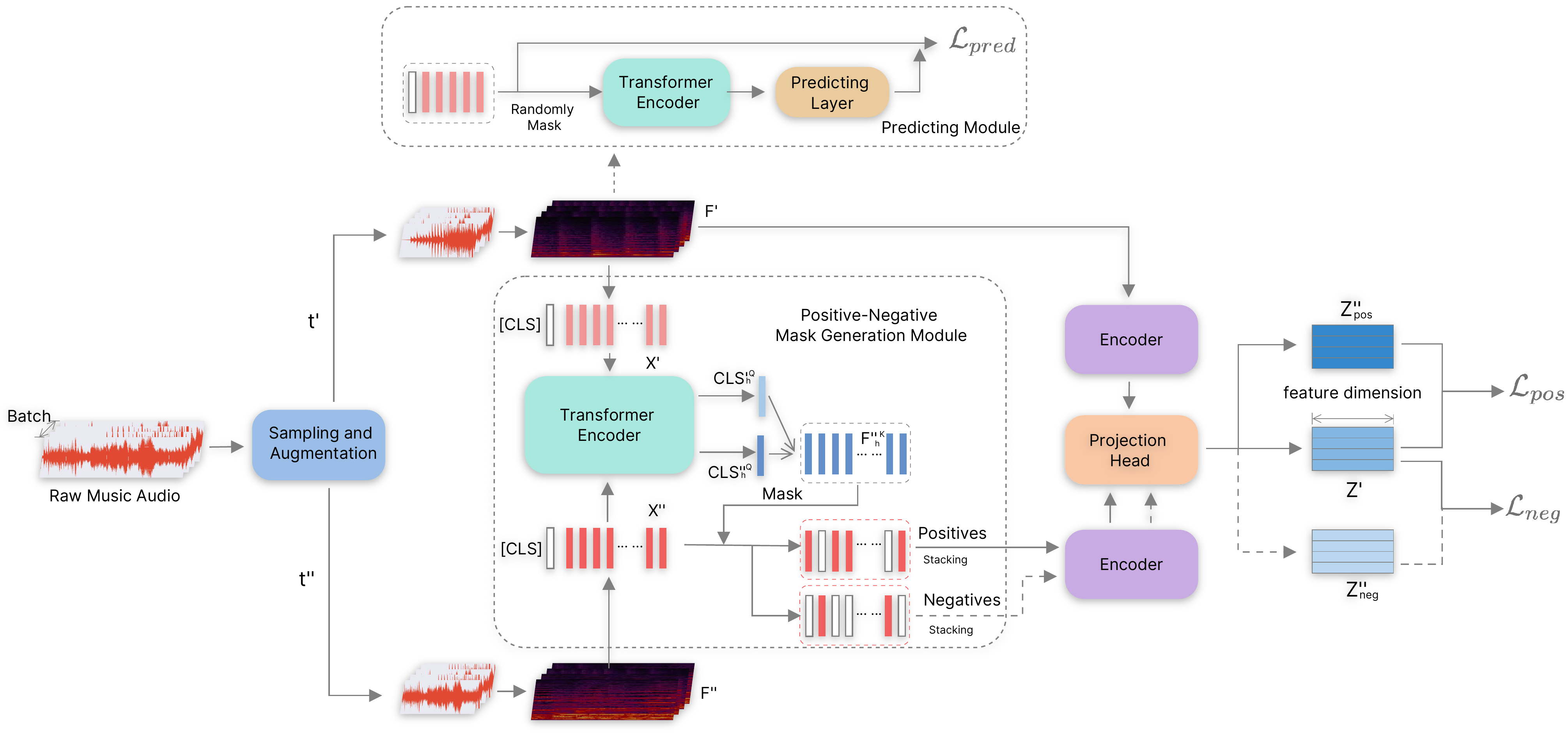}
		\caption{
			The overall framework of our proposed method for music representation learning. The predicting module captures the correlation among frames.
		}
		\label{fig:overview}
\end{center} \end{figure*} 
%--------------------------------fig end---------------------

The overall architecture of our pre-training framework is shown in Figure  \ref{fig:overview}. Our networks consist of a predicting module that utilizes a transformer encoder to learn contextual correlation among frames, an asymmetrical positive-negative mask generating module, and a contrastive learning module. After pre-training with music datasets,  we utilize the pre-trained \textbf{Encoder}, as previous works do \cite{Chen2020, He2020, Grill2020}, to obtain the general music representation for various downstream tasks, such as music classification and cover song identification.    
%For a raw audio signal of a music, we randomly crop two segments and apply augmentations to the clips from the audio signal for up and down branch respectively. 

\subsection{Sampling and Augmentation.} The function of this unit is to select two segments from the same waveform randomly and apply some augmentation methods in the selected segments. We use the same group of music augmentations as in CLMR \cite{Spijkervet2021}. The group includes polarity inversion, noise, gain, filter, delay, pitch shift. Each augmentation is randomly selected accordingly to its setting probability.

\subsection{Log-Mel Spectrogram}\label{logMel}

The digital audio signal represents that the voltage amplitude varies over time. According to the Fourier theorem, every signal can be decomposed into a set of cosine and sine waves that add up to the original signal, \textit{i.e.}, an audio signal is comprised of several single-frequency sound waves. We use its Mel spectrogram, generated by Short-Term Fourier Transformation (STFT) and Mel-scale filter banks to capture the time-domain and frequency-domain information of music raw waveform. The Mel-scale aims to mimic the human ear perception function--the human ear's sensitivity varies to different frequencies. In the deep learning domain, \cite{Conference2014} trained convolution networks to autonomously discover frequency decompositions from raw audio. For simplicity, we use STFT and Mel-scale filters to obtain the Mel spectrogram of music and convert it to a logarithmic scale, \textit{i.e.}, Log-Mel spectrogram.  

\subsection{Predicting Module}\label{trm}
Before generating masks, we should learn the correlation of each frame to the input music for more accurate masks. We use a random masking strategy in the input before feeding it to the transformer encoder. We then use a predicting layer to recover the masked positions from the output of the transformer encoder, obtaining a predicting loss. The Transformer encoder uses self-attention mechanisms primarily and learned or sinusoidal position information. Each layer consists of a self-attention sub-layer followed by a position-wise fully connected feed-forward network sub-layer.

Specifically, We view a frame as a token. In order to make the transformer encoder more stable and accurate when modeling the correlation among all tokens from a music fragment, we train it with a random mask strategy \cite{Devlin2019, Zhao2021, Liu2020}. we denote the frames set of a spectrogram as $\mathbf{F}=\left(\mathbf{f}_{1}, \mathbf{f}_{2}, \ldots, \mathbf{f}_{L}\right)$, where $\mathrm{F} \in \mathbb{R}^{L \times D}$. We append a learnable [CLS] token embedding in front of all frames, denoted as $\mathbf{X}=\left(\mathbf{c}, \mathbf{x}_{1}, \mathbf{x}_{2}, \ldots, \mathbf{x}_{L}\right)$, where $\mathrm{X} \in \mathbb{R}^{(L+1) \times D}$, so that we can aggregate information of all frames into [CLS] after attention operation. Then, we utilize a multi-head attention mechanism to calculate attention scores between a query and a key and use it for a value, which allows the model to focus on various parts of the frames sequence. The formulation of multi-head attention is,
\begin{align}
	\notag
	\mathbf{Y}_{n, h} = \operatorname{Attention}(\mathbf{Q}_{n,h}, \mathbf{K}_{n,h}, \mathbf{V}_{n,h}) \\
	=\operatorname{Softmax}\left(\frac{\mathbf{Q}_{n,h} \mathbf{K}_{n,h}^{T}}{\sqrt{D}}\right) \mathbf{V}_{n,h}
\end{align}
where $\mathbf{Q}_{n,h}$, $\mathbf{K}_{n,h} $, $\mathbf{V}_{n,h}$ are the query, key and value respectively. The \textit{n} and the \textit{h} are the index of layer and attention head respectively. They are calculated by $\mathbf{Q}_{n, h} = \mathbf{X}\mathbf{W}_{n, h}^{Q}$, $\mathbf{K}_{n,h} = \mathbf{X}\mathbf{W}_{n,h}^{K}$ and $\mathbf{V}_{n,h} = \mathbf{X}\mathbf{W}_{n,h}^{V}$. The $\mathbf{W}_{n,h}^{Q}$, $\mathbf{W}_{n,h}^{K}$ and $\mathbf{W}_{n,h}^{V}$ $\in \mathbb{R}^{D \times D}$, which are the corresponding weight matrices. The attention scores between $\mathbf{Q}_{n,h}$ and $\mathbf{K}_{n,h}$ are divided by $\sqrt{D}$ to avoid large values of the dot product. Because the self-attention can not aware of the order, we add a sinusoidal position embedding to the frames sequence before input to self-attention.

The multi-head attention aggregates contextual information through learnable weights, but it is still a linear model. For introducing the non-linearity, multi-head attention output will be fed to the position-wise feed-forward network (FFN). Specifically, within the \textit{n}th layer of transformer encoder, we concatenate the outputs of all attention heads and apply linear transformation to get $\mathbf{Y}_{n}$, and input it into FFN to obtain output $\mathbf{X}_{n}$,
\begin{align}
	\mathbf{Y}_{n} = \operatorname{Concat}(\mathbf{Y}_{n, 1}, \mathbf{Y}_{n, 2}, \ldots, \mathbf{Y}_{n, h}, \ldots, \mathbf{Y}_{n, H})\mathbf{W}_{n}
\end{align}
\begin{align}
	\mathbf{X}_{n} = \operatorname{ReLu}(\mathbf{Y}_{n}\mathbf{W}_{n,1}+\mathbf{b}_{n,1})\mathbf{W}_{n,2}+\mathbf{b}_{n, 2}
\end{align}
where $\mathbf{W}_{n} \in \mathbb{R}^{HD \times D}$, $\mathbf{W}_{n,1}$ and  $\mathbf{W}_{n,2} \in \mathbb{R}^{D \times D}$, $\mathbf{b}_{n,1}$ and $\mathbf{b}_{n,2} \in \mathbb{R}^{D}$. The \textit{H} is the number of attention heads. We randomly mask the frames and input into the above transformer encoder. And we then use FFN to predict the masked content of the output of transformer encoder so that we can learn robust contextual information between frames. The output of transformer encoder here is denoted as $\mathbf{P}$. For other details about transformer, such as positional encoding and residual connections, you can find from \cite{Vaswani2017}. The protocol of random masking follows \cite{Devlin2019}.

\subsection{Generating Positive-Negative Frame Mask}

Not all frames of a piece of music play an equivalent role in characterizing songs. Randomly masking the frames of music is a straightforward method to avoid the adverse effect of the inconsequential music frames. However, that may mask the critical frames of music, resulting in learning the inaccurate representation of the music. Simultaneously, the trivial frames will be retained and encoded, which will deteriorate the learned representation to a certain extent. Therefore, it is indispensable to approximately quantify the importance of a single frame to the entire music to mask frames selectively. We argue that the crucial frames can facilitate us learning a more distinct music representation that the network can identify different music more efficiently and accurately. As we describe above, the noisy or inessential parts will be detrimental to the music representation. Reducing the effect of non-critical frames is necessary. Hence, we generate the positive and negative masks to create the augmented positive and counterfactual negative representation. The contrastive learning loss functions will optimize the agreement between positives the distance between positives and negatives. More details can be seen in Section \ref{loss}. 

We design an asymmetrical module to obtain the positive-negative mask. Firstly, we randomly select two fragments from the same music raw waveform. After applying several augmentation approaches, we get their Log-Mel spectrogram produced by stacking many frames. Before inputting them into the transformer encoder, we add a [CLS] token vector in front of their frames sequences. As we illustrate above, the vector $\mathbf{c}$ will contain the information of all frames within a sequence after being encoded by the transformation encoder. The first token of two branches will be $\mathbf{c}^{'}$ and $\mathbf{c}^{''}$ respectively. Therefore, we use the query vector of $\mathbf{c}^{'}$ and $\mathbf{c}^{''}$ to calculate attention scores between it and the keys of frames $\mathbf{F}^{''}$. The attention scores will be used to select a certain percentage of frames to be masked. Specifically, within the last layer of transformer encoder, we take out the $\mathbf{CLS}_{h}^{'Q}$ and $\mathbf{CLS}_{h}^{''Q}$ of added tokens in both branches. Then, we utilize these queries to calculate the attention scores with $\mathbf{F}_{h}^{''K}$,
\begin{equation}
\begin{split}
\mathbf{s} = \frac{1}{2}\sum_{h=1}^{H} (\operatorname{Softmax}(\mathbf{CLS}_{h}^{'Q} \cdot \frac{\mathbf{F}_{h}^{''K}}{\sqrt{D}}\\
+ \operatorname{Softmax}(\mathbf{CLS}_{h}^{''Q} \cdot \frac{\mathbf{F}_{h}^{''K}}{\sqrt{D}}))
\label{scores}
\end{split}
\end{equation}
where $\mathbf{F}_{h}^{''K}$ is the keys in the \textit{h}th attention head, $\mathbf{CLS}_{h}^{'Q}$ and $\mathbf{CLS}_{h}^{''Q}$ $\in \mathbb{R}^{1 \times D}$,  $\mathbf{F}_{h}^{''K} \in \mathbb{R}^{D \times L}$, $\mathbf{s}  \in \mathbb{R}^{1 \times L}$. The frames with high values in the $\mathbf{s}$ mean that they are crucial to both music fragments. According to the $\mathbf{s} $, we can screen a certain proportion of the frames with the lower attention weights. The remained crucial frames will reserve the global and local information since the two segments locate in different positions of the whole track and they are critical to the two segments. Specifically, we rank the $\mathbf{s} $ in the ascending order and set the value ranked at ratio $\boldsymbol{r}$ as the threshold $\boldsymbol{t}$. The $\boldsymbol{r}$ is the ratio value, and we set it to 10\% as the default value. We obtain the positive mask matrix $\mathbf{M} = (\mathbf{m}_{1}, \mathbf{m}_{2}, \ldots, \mathbf{m}_{i}, \ldots, \mathbf{m}_{L})$ as follows,
\begin{equation}
\mathbf{m}_{i} = \begin{cases}\mathbf{0}, & \mathbf{s}_{i} <   \boldsymbol{t} \\
\mathbf{e}, & o t h e r s \end{cases}
\end{equation}
where $\mathbf{e}$ is unit vector, $\mathbf{0}$ is a zero vector. The negative mask $\overline{\mathbf{M}} = 1 - \mathbf{M}$. We add the positive and negative mask to the input frames $\mathbf{F}^{''}$ to obtain the augmented positive frames $\mathbf{F}_{pos}^{''}$ and counterfactual negative frames $\mathbf{F}_{neg}^{''}$ respectively. The $\mathbf{F}_{pos}^{''}$ and $\mathbf{F}_{neg}^{''}$ will be encoded and projected into the positive representation $\mathbf{Z}^{''}_{pos}$ and negative representation $\mathbf{Z}^{''}_{neg}$. 
%s d $\mathcal{L}_\mathcal{B T}$
% $\mathcal{L}_\mathcal{CONS}$

% %--------------------------------fig begin---------------------
% \begin{figure}[t] \begin{center}
%     \includegraphics[width=\columnwidth]{figures/Mask Generator.pdf}
%     \caption{
%      We randomly mask frames to train transformer encoder and utilize its Query ($\mathbf{Q}$) and Keys ($\mathbf{K}$) to steer generating Frame-mask.   
% 	}
% \label{fig:maskGen}
% \end{center} \end{figure} 
% %--------------------------------fig end---------------------

\subsection{Encoder and Projection Head}\label{encoder}

As the familiar setting in contrastive learning \cite{Chen2020, Grill2020, Spijkervet2021}, we apply a neural network encoder $f(\cdot)$ to extract representation vectors from augmented examples and use a MLP with one hidden as projection head $g(\cdot)$ to map representation to the latent space where contrastive loss is applied. We opt Fully Convolutional Networks (FCN) \cite{Choi2016} as our base encoder. The dimensionality of representation vectors from encoder is $D_{e}$ = 512, from projection head is $D_{p}$ = 256.

\subsection{Pre-training Objective Function}
\label{loss}

For training the model, we adopt Huber loss \cite{Girshick2015}, and Barlow Twins loss \cite{Zbontar2021} as our pre-training objective. In section \ref{trm}, a predicting layer is to predict the random disturbed input according to the output of the transformer encoder. The set $\mathbf{I}$ includes all masked frames' index. We calculate the predicting loss $\mathcal{L}_{pred}$ as follows,
\begin{equation}
\mathcal{L}_{pred} = \sum_{i \in \mathbf{I}} \sum_{j=0}^{D} \operatorname{smooth}_{L_{1}}(\mathbf{X}_{i,j} - \mathbf{P}_{i,j}),
\end{equation}

\begin{equation}
\label{smooth}
\operatorname{smooth}_{L_{1}}(x) =  \begin{cases}0.5 \cdot x^{2}, & |x|<1 \\ |x|-0.5 , & otherwise \end{cases}
\end{equation}

L2 loss is more sensitive to outliers due to the square function. To stabilize training, we follow \cite{Girshick2015} to use L1 loss when |x| is larger than 1 as equation \ref{smooth}. So the $\mathcal{L}_{pred}$ is less sensitive to outliers. In section \ref{encoder}, we feed a batch of $\mathbf{F}^{'}$, the augmented positive version and counterfactual negative version of a batch of $\mathbf{F}^{''}$ into encoder and projection head to respectively get $\mathbf{Z}^{'}$, $\mathbf{Z}^{''}_{pos}$, $\mathbf{Z}_{neg}^{''} \in \mathbb{R}^{B \times D_{p}}$, where B is the value of batch size. The $\mathbf{Z}^{'}$ and $\mathbf{Z}^{''}_{pos}$ are treated as the positive samples in the contrastive space while $\mathbf{Z}^{''}_{neg}$ is the negative samples. We compute the contrastive loss between $\mathbf{Z}^{'}$ and $\mathbf{Z}^{''}_{pos}$, denoted as $\mathcal{L}_{pos}$ in the following manner,
\begin{equation}
\mathcal{L}_{pos} = \sum_{i=0}^{D_{p}}(1 - \mathbf{U}_{i,i})^{2} + \lambda \sum_{i=0}^{D_{p}} \sum_{j \neq i}^{D_{p}} \mathbf{U}_{i, j}^{2}
\label{cont}
\end{equation}
where $\mathbf{U} \in \mathbb{R}^{D_{p} \times D_{p}}$  is the cross-correlation matrix between $\mathbf{Z}^{'}$ and $\mathbf{Z}_{pos}^{''}$. $\mathcal{L}_{pos}$ is the same as BTLoss \cite{Zbontar2021}. Meanwhile, we design a contrastive loss for the negative samples,
\begin{equation}
\mathcal{L}_{neg} = \lambda \sum_{i=0}^{D_{p}} \mathbf{V}_{i,i}^{2}\label{cont}
\end{equation}
where $\mathbf{V} \in \mathbb{R}^{D_{p} \times D_{p}}$  is the cross-correlation matrix between $\mathbf{Z}^{'}$ and $\mathbf{Z}_{neg}^{''}$. The $\lambda$ is a hyperparameter to trade off the importance of $\mathcal{L}_{neg}$ and the second term of $\mathcal{L}_{pos}$. The loss $\mathcal{L}_{pos}$ and $\mathcal{L}_{neg}$ contrast data samples along the feature dimension which can prevent trivial constant solutions. Our final loss is $\mathcal{L}$ =  $\mathcal{L}_{pred}$ + $\mathcal{L}_{pos}$  + $\mathcal{L}_{neg}$.
\section{Experimental Evaluation}
% Transfer learning is a machine learning approach where a model developed for a task is reused as a starting point for a model for a second task. This is a popular approach in deep learning, and given the considerable computational cost and time resources required to develop neural network models, the pre-trained model is used as a starting point for computer vision and natural language processing tasks. . We do transfer experiments in two tasks, music classification, and cover song identification.
The primary purpose of unsupervised learning is to learn generalized and transferrable representation. Therefore, it is necessary for us to testify that if our learned music representation owns outstanding generalization and transferability. Following commonly used evaluation protocol of pre-trained representation \cite{Chen2020, He2020}, we firstly do linear evaluation, semi-supervised learning, transfer learning for music classification. We then transfer the learned representation to another dataset of cover song identification.

\subsection{Experimental Setting}

\subsubsection{Dataset.} We experiment with several available public datasets ofter used for classification and cover song identification. More details about datasets are illustrated as follows,

\begin{itemize}[leftmargin=*]
	\item \textbf{MagnaTagATune (MTAT):} The annotations of MagnaTagATune were collected by Edith Law’s TagATune game \cite{Law2009}. The dataset includes 25863 pieces of music which are 29-seconds-long, and each track has multiple tags. The clips span a broad range of genres like Classical, New Age, Electronica, Rock, Pop, World, Jazz, Blues, Metal, Punk, and more. We split it into train/valid/test with a ratio as \cite{Spijkervet2021}, and get 18706/1825/5329 tracks; 
	
	\item \textbf{GTZAN\footnote{http://marsyas.info/downloads/datasets.html} : } The dataset consists of approximately 1000 audio tracks, each 30 seconds long. It contains ten genres.
	
	\item \textbf{Second Hand Songs 100K (SHS100K):} \textbf{A cover version is a new performance or recording by a musician other than the original performer of the song.} We crawled raw audios through youtube-dl\footnote{https://github.com/ytdl-org/youtube-dl} using the provided URLs from Github\footnote{https://github.com/NovaFrost/SHS100K2}. Due to the copyright, We crawled all songs from YouTube and got 9733 songs. Every song has many cover versions. All cover versions of the 9733 songs add up to 104612. Following the experimental setting of \cite{Yu2019}, we selected the songs whose number of cover song versions were larger than 5 for training. We randomly selected tracks from the remaining records to construct two subsets for validation and testing, respectively. The ratio among training set, validation set, and testing set is 8:1:1. We get 6000 songs with 84153 versions for training, 1941 songs with their 10456 cover songs for testing; 
	
	\item \textbf{Covers80\footnote{https://labrosa.ee.columbia.edu/projects/coversongs/covers80/}:} There are 80 songs exists in Covers80, and every song has 2 cover versions.
\end{itemize}

%--------------------------table-----------------------------------
\begin{table}[t]
	\centering
	\caption{The statistics of all datasets.}
	\begin{tabular}{c c c c}
		\toprule
		Dataset & train & validation & test \\
		\midrule
		MagnaTagATune & 18,706 & 1,825 & 5,329 \\
		GTZAN         & 930 & - & -  \\
		SHS100K       & 9,999 & - & 1,004  \\
		Covers80      & - & - & 160  \\
		\bottomrule
	\end{tabular}
	\label{dataset}
\end{table}
%--------------------------table-----------------------------------

\subsubsection{Metrics.} To evaluate our learned representation for music, we follow the commonly used linear evaluation setting \cite{Kolesnikov2019, Chen2020, Spijkervet2021, Bachman2019}, where a linear classifier is trained on a pre-trained encoder from which parameters are not updated. Moreover, we train a multi-layer perceptron (MLP) to observe if the performance can be better after adding the depth of the classifier. We choose ROC-AUC and PR-AUC to measure the effect of the classifier comprehensively. For the cover song identification task, we use the widely used evaluation metrics \footnote{https://www.musicir.org/mirex/wiki/2020:Audio\_Cover\_Song\_Identification} mean average precision (MAP), the mean rank of the first correctly identified cover (MR1), and precision at 10 (Precision@10). Precision@10 is the mean ratio of the identical versions recognized successfully in the top 10 ranking list, which is obtained through ranking all records by the similarity between query and references. We calculate scalar products between two music representation to judge their similarity.

\subsubsection{Implementation Details} 
The transformer encoder in the Predicting Module shares the parameters with the transformer encoder in the Positive-Negative Mask Generation Module. The basic encoders, a full convolution network with 4 layers \cite{Choi2016}, share parameters between two branches. The encoder outputs a 512-dimension feature as a representation. An MLP as projection head is used to map the representation to a smaller space. The output dimension of the projection head is 256. It is worth mentioning that the encoders and projection heads of all the unsupervised methods used in the experiments are consistent. We use the Adam optimizer. The learning rate is 0.0003, and weight decay is 1.0 $\times 10^{-6}$. Others are default. We set the batch size to 48 and train for 300 epochs, taking about 30 hours on a GPU. At the spectrogram extracting stage, the hop size is 128 during time-frequency transformation. STFT is performed using 256-point FFT while the number of mel-bands is set as 128. The transformer encoder consists of 3 layers, and its multi-head attention sub-layer has 3 heads.

%--------------------------table-----------------------------------
\begin{table}[t]
	\centering
	\caption{The performance of some advanced supervised and self-supervised methods in music classification tasks is all trained on the MagnaTagATune dataset. For the unsupervised models, the scores are obtained by linear classifiers. * represents the performance of an MLP classifier.}
	
	\resizebox{\columnwidth}{!}{    
		\begin{tabular}{c c c c c}
			\toprule
			& Method & Param & ROC-AUC & PR-AUC \\
			\midrule
			\multirow{4}{*}{Supervised} 
			& 1D-CNN & 382K & 85.6 & 29.6 \\
			& SampleCNN & 2394K & 88.6 & 34.4 \\
			& Musicnn  & 228K & 89.1 & 34.9 \\
			& Timber CNN & 220K & 89.3 & - \\
			& FCN-4     & 370K & 89.4 & - \\
			\midrule
			\multirow{5}{*}{Self-Supervised} 
			& MoCo & 370K & 87.0 & 32.1 \\
			& MoCo v2 & 370K  & 87.9 & 33.2 \\
			& CLMR & 2394K & 88.5 & 35.4 \\
			& SimCLR & 370K & 88.7  & 34.8  \\
			& BYOL & 370K & 89.1 & 35.8 \\
			& PEMR(ours) & 370K & \textbf{89.6} & \textbf{36.9} \\
			\midrule
			\multirow{5}{*}{Self-Supervised} 
			& CLMR* & 2394K & 89.3 & 35.9 \\
			& MoCo* & 370K & 89.5 & 36.3 \\
			& MoCo v2* & 370K & 89.8 & 36.6 \\
			& SimCLR* & 370K & 89.8 & 36.9 \\
			& BYOL* & 370K & 89.9 & 37.0 \\
			& PEMR(ours)* & 370K & \textbf{90.3} & \textbf{38.0} \\
			
			\bottomrule
	\end{tabular}}
	
	\label{modPer}
\end{table}
%--------------------------table-----------------------------------

\subsection{Music Classification}

We select some traditional and advanced supervised baselines in music classification and select some state-of-the-art self-supervised baselines. To make the results more persuasive, we implement several advanced contrastive learning models for music representation. 

\subsubsection{Linear Evaluation.}

Table \ref{modPer} shows the performance comparison in music classification task among many approaches, including supervised methods and self-supervised methods. We follow \cite{Chen2020, He2020} to calculate the number of parameters of encoders in self-supervised methods. We use the thop package\footnote{https://pypi.org/project/thop/} to obtain the model size. Following a standard linear evaluation setting \cite{Chen2020, Grill2020, He2020}, We use the training set of MTAT to pre-train the models. After that, we train linear classifiers based on the frozen pre-trained encoder to evaluate the test dataset of MTAT. The applied encoders are the same for all methods except CLMR. CLMR uses sampleCNN \cite{Lee2019} adopting raw waveform as model input. To ensure a faithful comparison, the metric values of baselines are directly copied from their papers, where Timber CNN and FCN-4 do not report PR-AUC values. Our method achieves the best performance under linear evaluation protocol. We attribute the empirical results to the positive-negative frame mask, making the networks preserve the context while concentrating on the critical parts of music. The comparison between CLMR and other self-supervised methods demonstrates the advantage of the Log-Mel spectrogram. In addition, we train the MLP to observe whether the performance can be improved when introducing more parameters. The experimental results of PEMR reach the best of 90.3\% in ROC-AUC, 37.8\% in PR-AUC.

\subsubsection{Semi-Supervised Learning.}

Getting tagged data for deep learning problems often requires skilled human agents. As a result, the costs associated with the labeling process can make a large number of fully labeled training sets infeasible while obtaining unlabeled data is relatively inexpensive. In such situations, semi-supervised learning can be of great practical value. Aiming at estimating if our learned music representation can still perform well in the semi-supervised learning classification task, we decrease the percentage of the labeled training data during the fine-tuning stage. Specifically, we randomly sample 1\%, 10\% labeled data from MTAT training dataset just as \cite{Beyer2019, Chen2020} do. We feed these few labeled data directly to the pre-trained base encoders and linear classifiers for training. The evaluation results of the previous approaches and PEMR are shown in Table \ref{semiPer}. In contrast with other musical self-supervised methods, PEMR can generate more generalized music representation even if labeled music for learning is inadequate. Besides, we randomly initialize the base encoder FCN carrying a linear classifier and train it with the same sampled labeled data. Our performance substantially exceeds the model trained from scratch. The empirical results prove the significance of our pre-trained music representation at a labeled-data-lacked scenario.

%--------------------------table-----------------------------------
\begin{table}[t]
	\centering
	\caption{We fine-tune the pre-trained encoders and linear classifiers with different quantities of labeled data.}
	\resizebox{\columnwidth}{!}{ 
		\begin{tabular}{c c c c c}
			\toprule
			\multirow{3}{*}{Method} & \multicolumn{4}{c}{Label Fraction} \\
			& \multicolumn{2}{c}{1\%} & \multicolumn{2}{c}{10\%} \\
			\cmidrule(lr){2-3} \cmidrule(lr){4-5}
			& ROC-AUC & PR-AUC & ROC-AUC & PR-AUC \\
			\midrule
			FCN & 73.2 & 19.7 & 86.3 & 30.8 \\ 
			\midrule
			MoCo & 74.3 & 18.6 & 87.4 & 33.1 \\
			MoCo v2 & 75.3 & 20.1 & 87.2 & 32.4 \\
			CLMR & 77.3 & 22.6 & 87.0 & 32.9 \\
			SimCLR & 73.1 & 18.6 & 87.2 & 32.5 \\
			BYOL & 76.2 & 22.4 & 87.6 & 33.4 \\
			
			PEMR(ours) & \textbf{77.3} & \textbf{24.4} & \textbf{88.0} & \textbf{34.0} \\
			
			\bottomrule
	\end{tabular}}
	\label{semiPer}
	
\end{table}
%--------------------------table-----------------------------------

\subsubsection{Transfer Learning}

%--------------------------table-----------------------------------
\begin{table}[t]
	\centering
	\caption{We pre-train the models in the GTZAN dataset and transfer their learned parameters to another dataset for training. We evaluate the transfer capacity in both linear evaluation and fine-tuning settings.}
	\begin{tabular}{c c c c c}
		\toprule
		\multirow{2}{*}{Method} & \multicolumn{2}{c}{Linear Evaluation} & \multicolumn{2}{c}{Fine-tuned} \\
		\cmidrule(lr){2-3} \cmidrule(lr){4-5}
		& ROC-AUC & PR-AUC & ROC-AUC & PR-AUC \\
		\midrule
		FCN  & - & - & 89.8 & 36.8 \\
		\midrule
		MoCo  & 74.8 & 18.9 & 89.6 & 36.6  \\
		MoCo v2 & 78.4 & 21.3 & 89.8 & 36.8 \\
		CLMR  & 81.9 & 26.2 & 89.7 & 36.1 \\
		BYOL  & 86.7 & 32.0 & 89.3 & 36.0 \\
		SimCLR   & 86.8 & 32.1 & 89.6 & 36.6 \\
		PEMR(ours)  & \textbf{87.9} & \textbf{33.8} & \textbf{90.0} & \textbf{37.3}  \\
		\bottomrule
	\end{tabular}
	\label{tab:transCLS}
\end{table}
%--------------------------table-----------------------------------

Specifically, we adopt the whole GTZAN dataset for pre-training the model and employ the MTAT training dataset to train classifiers that will evaluate the MTAT testing dataset. To reveal the superiority of PEMR more apparently, we compare several self-supervised methods for music representation, including previous approaches and ours. As shown in Figure \ref{tab:transCLS}, the effect of the classifiers based on the encoder pre-trained by PEMR is the most advanced under both linear evaluation and fine-tuning settings. More importantly, we can surpass the same network trained in the supervised learning paradigm in the fine-tuning situation. Furthermore, although we freeze the pre-trained encoder when training the classifier, we can also obtain competitive results of 87.9\% in ROC-AUC and 33.8\% in PR-AUC. The results are comparable to the supervised FCN trained from scratch.

\subsection{Cover Song Identification}

The task of cover song identification is to identify alternative versions of previous musical works. Since different versions of a song are performed by various artists or musicians and instruments, they may vary significantly in pitch, rhythm, structure, and even in fundamental aspects related to the harmony and melody of the song. Recently, cover song recognition has attracted attention because it has the potential to serve as a benchmark for other musical similarities and retrieval algorithms. Chord analysis, melodic extraction, and musical similarity are all closely related to cover song identification -- another area of music analysis where artificial intelligence is used. Before \cite{Xu2018}, previous research primarily involved hand-crafted features, which was intolerable when facing large-scale datasets. Given this, \cite{Xu2018} proposed deep learning methods, learning to extract features efficiently for cover song identification. \cite{Yu2019} and \cite{Yu2020} devised TPP-Net and CQT-Net that could be naturally adapted to deal with key transposition in cover songs and they designed a training scheme to make their model more robust. We select these advanced models as our baselines of the cover song identification task. The main goal of them is to learn the high-quality representation of songs, employing supervised methods. There is still a great developing space for applying self-supervised learning in this task.

The pre-training for music representation will be greatly meaningful if the pre-trained music representation can be transferred to other downstream tasks where training datasets have little labeled data. After pre-training in the MTAT training dataset, we obtain the encoder and fine-tune it with the datasets from the cover song identification domain. The more details are as follows: 1) the network we want to train consists of FCN and CQT-Net. 2) the SHS100K training set is provided for the network to fine-tune. 3) we extract music representation through the trained network and evaluate their performance on the SHS100K testing set and Covers80 dataset. It is worth mentioning that \textbf{many songs can not be downloaded from YouTube due to the invalid copyright}, resulting in the much difference between our downloaded SHS100K and SHS100K used in previous methods. So we randomly sample the data from SHS100K to construct a subset of SHS100K, namely SHS100K-SUB, and split it into train, validation, test set with the same ratio as \cite{Yu2019, Yu2020}. Table \ref{tab:coverPer} exhibits our experimental results. We randomly initialize the parameters of FCN and CQT-Net. After training in the SHS100K-SUB train set, performance in the SHS100K-SUB test set or Covers80 can not surpass the individual CQT-Net. Nevertheless, we can improve the model's performance on two datasets by incorporating the pre-trained FCN with CQT-Net. 

%--------------------------table-----------------------------------
\begin{table}[t]
	\centering
	\caption{Transfer learning for cover song identification. The music representation encoder used in this task is pre-trained with the out-of-domain dataset.}
	\begin{tabular} {c c c c c c c}
		\toprule
		Method & & & & MAP & Precision@10 & MR1 \\
		\midrule
		& & & & \multicolumn{3}{c}{SHS100K-SUB} \\
		\midrule
		Ki-Net & & & & 0.112 & 0.156 & 68.33\\
		TPP-Net & & & & 0.267 & 0.217 & 35.75 \\
		FCN     & & & & 0.289 & 0.230 & 34.86 \\
		CQT-Net & & & & 0.446 & 0.323 & \textbf{18.09} \\
		\midrule
		\multicolumn{6}{l}{\textit{Fine-tuned:}} \\
		(rand. FCN) + CQT-Net & & & & 0.433 & 0.317 & 21.13 \\
		(pre. FCN) + CQT-Net & & & & \textbf{0.484} & \textbf{0.341} & 20.68 \\
		\midrule
		& & & \multicolumn{3}{c}{Covers80} \\
		\midrule
		Ki-Net & & & & 0.368 & 0.052 & 32.10\\
		TPP-Net & & & & 0.5 & 0.068 & 17.08 \\
		FCN     & & & & 0.529 & 0.073 & 12.50 \\
		CQT-Net & & & & 0.666 & 0.077 & 12.20 \\
		\midrule
		\multicolumn{6}{l}{\textit{Fine-tuned:}} \\
		(rand. FCN) + CQT-Net & & & & 0.624 & 0.079 & 14.43 \\
		(pre. FCN) + CQT-Net  & & & & \textbf{0.668} & \textbf{0.081} & \textbf{10.52} \\
		\bottomrule
	\end{tabular}
	\label{tab:coverPer}
	
\end{table}
%--------------------------table-----------------------------------

\subsection{Ablation Study}

\begin{figure}[t]
	
	\centering
	\includegraphics[width=\linewidth]{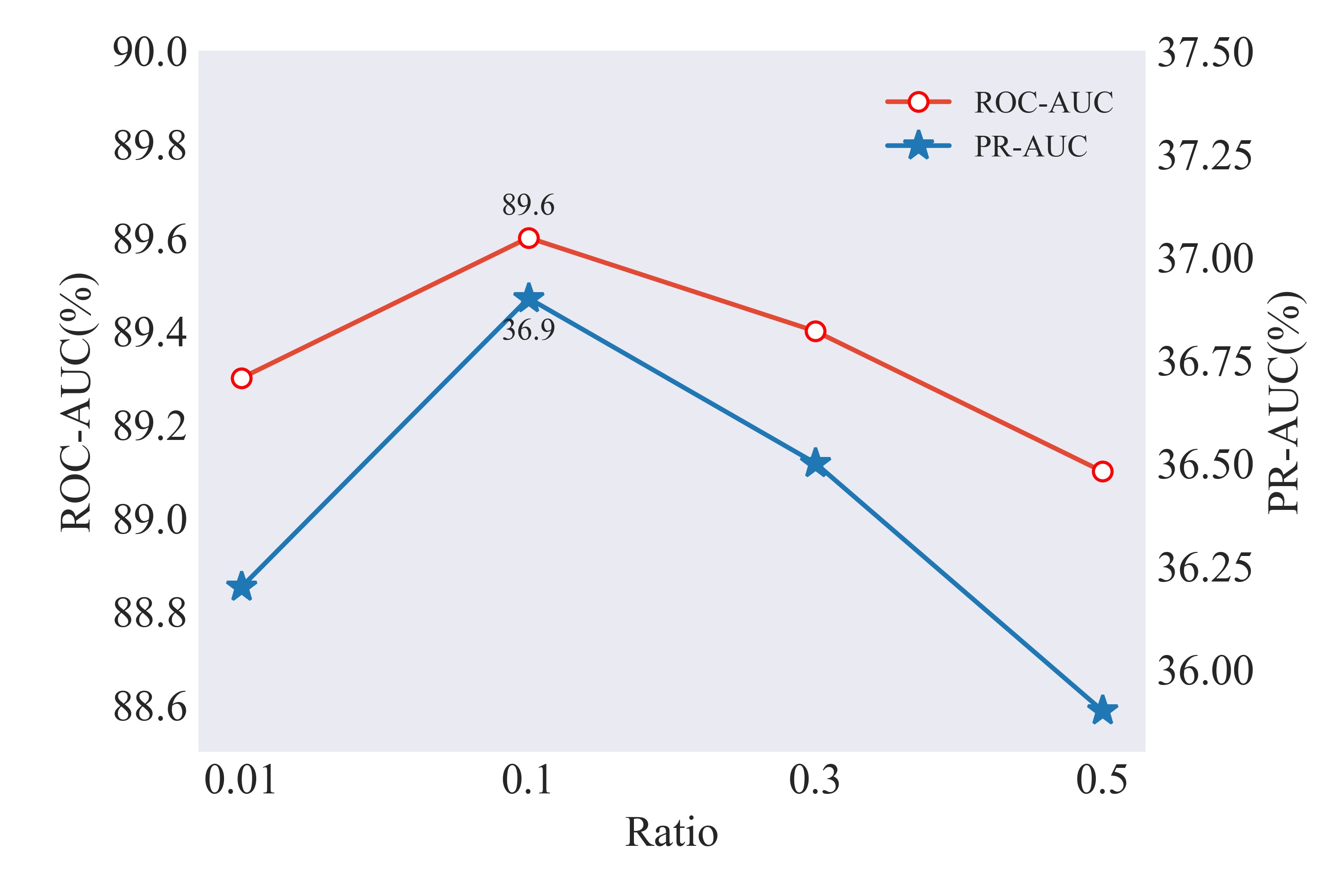}
	%\caption{fig2}
	
	\caption{
		The variation of the performance with different masked ratios.
	}
	\label{ratio}
\end{figure}

This section will analyze the impact of the augmented positive representation and the counterfactual negative representation generated from the positive-negative mask on learning high-quality music representation; Besides, We modify the vital parameter \textit{ratio} when creating the positive-negative mask to control the proportion of the masked frames in input data. Specifically, the ratio $\boldsymbol{r} \in$ [0.01, 0.1, 0.3, 0.5], we change its value and pre-train the network from scratch for 300 epochs. The experimental results are shown in Table \ref{ablation} and plotted in Figure \ref{ratio},respectively. We find that,
\begin{itemize}[leftmargin=*]
	
	\item We pre-train the baseline model without any masking strategy; Based on the baseline, we only add a positive mask into the model. Their experimental results are shown in Table \ref{ablation}. Generating the augmented positive frames and the counterfactual negative frames is beneficial for the model to learn effective music representation.
	
	\item In Figure \ref{ratio}, the model achieves the best result when $\boldsymbol{r}$ is 0.1 while suffering from performance drops if the $\boldsymbol{r}$ continues growing. We ascribe this phenomenon to: 1) the mild effect of the mask. When $\boldsymbol{r}$ is too small, for example, $\boldsymbol{r}$ equals to 0.01, which means little low-scores frames selected to construct the negative frames sequence $\mathbf{F}_{neg}^{''}$. Most of the noisy or inessential frames will be retained in the positive frames sequence $\mathbf{F}_{pos}^{''}$, resulting that the model can not focus on the crucial elements. 2) the excessive effect of the mask. The negative frames sequence will contain a large number of the crucial frames if $\boldsymbol{r}$ is too large. That will destroy the positive frames sequence, causing the model to learn inaccurate music representation.
	
\end{itemize}

%--------------------------table-----------------------------------
\begin{table}[t]
	\centering
	\caption{We experiment with a baseline contrastive framework without the mask. Then, we apply the positive mask and the negative mask in order. }
	\begin{tabular}{c c c c c}
		\toprule
		Method & & ROC-AUC & & PR-AUC \\
		\midrule
		w.o. mask & & 89.1 & & 36.2 \\
		pos. mask & & 89.4 & & 36.6 \\
		pos.+ neg. mask (PEMR) & & 89.6 & & 36.9 \\
		\bottomrule
	\end{tabular}
	\label{ablation}
\end{table}
%--------------------------table-----------------------------------

%--------------------------------fig-------------------------
%--------------------------------fig end--------------------

% \subsubsection{Summary of Transfer Learning}

% The experimental results of transfer learning in Table \ref{tab:transCLS} and Table \ref{tab:coverPer} can both demonstrate our positive-negative frame mask is significant for the model to learn effective music representation. The positive-negative mask will be produced by the designed asymmetrical module. More concretely, the positive mask can shield the inessential elements, thus obtaining the augmented positive frames sequence, encoded and projected as the augmented positive representation in the contrastive space. The augmented positive representation contains the global and local crucial information since the augmented positive frames are critical for the two sampled music fragments having different locations in the complete music. On the contrary, the negative mask will sift the global and local inessential information to be encoded in the counterfactual negative representation. Pushing the positive representation away from the negative representation can make the model pay more attention to the crucial region of the music, thus capturing the key features of the music to generate the more remarkable and effective music representation.

\begin{figure}[h]
	\centering
	\includegraphics[width=\columnwidth]{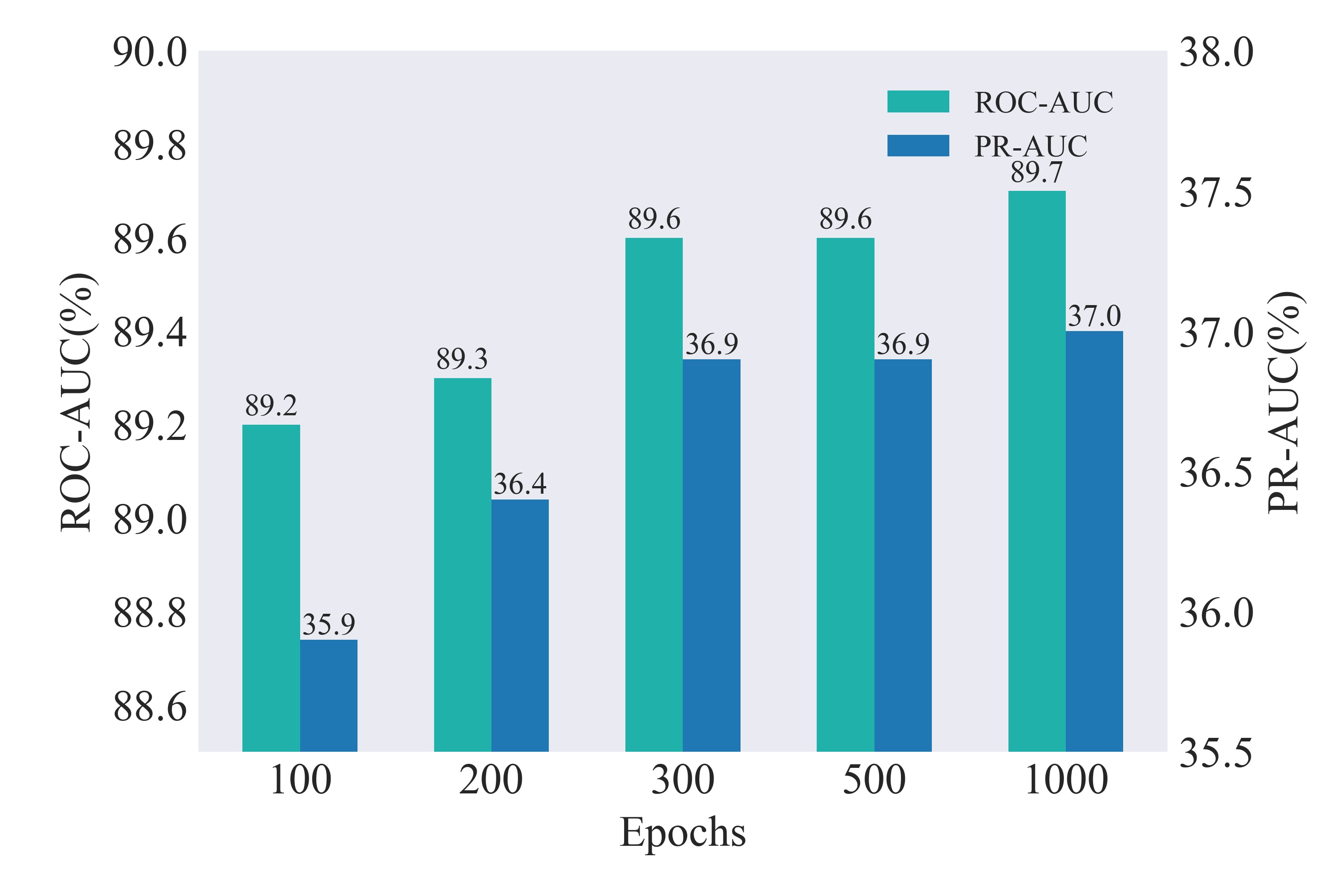}
	\caption{Linear music classifier trained on the top of our pre-trained encoder pre-trained with different epochs.}
	\label{epochs}
\end{figure}

\subsection{Training Epochs}

Figure \ref{epochs} shows the impact of different numbers of training epochs. When training time is relatively short, the training epoch is a critical key influencing the final performance. With more training epochs, the gaps between different epochs decrease or disappear. 

\section{Conclusion and Future Work}

In this paper, we propose to mask the critical and unimportant or noisy regions of music under the contrastive learning framework so that the model can concentrate on the crucial parts of the music, thus learning the more remarkable and effective representation for music. We devise an asymmetrical module to obtain the positive-negative mask by utilizing the transformation encoder's attention weights. Our pre-trained representation is applied to music-related downstream tasks: music classification and cover song identification. The experimental results of two music-related tasks demonstrate that the positive-negative mask is beneficial for the model to learn more effective music representation, which has strong generalization ability and transferability. 

However, there are still many challenges in the Music Information Retrieval (MIR) community, such as music recommendation system, music source separation, instrument recognition, and music generation, including music classification and cover song identification. Applying pre-trained music representation to music-related areas is a potential way to solve these challenges. We look forward to more research work in this field. 
%   Using the pre-trained model can be one of the effective ways to solve the above challenge. 
% We can apply the pre-trained music representation to speed up training and improve performance. Currently, the work of contrastive learning used in the music domain is little. As far as w know, our work is the first attempt to employ the mask under the contrastive learning framework to generate augmented positive samples and negative samples for music representation. We look forward to more related work in the future.
\section{Acknowledgments}
The work is supported by the Zhejiang Natural
Science Foundation (LR19F020006), and National Natural Science Foundation
of China (No. 61836002,62072397,62037001)

\clearpage
\balance
\bibliographystyle{ACM-Reference-Format}
\bibliography{citiation.bib}

\end{document}